\documentclass[epj]{svjour}
\usepackage{graphics}
\usepackage{lineno,hyperref}
\usepackage{graphicx}
\usepackage{url}
\usepackage{float}
\usepackage{latexsym}
\usepackage{amsmath,amssymb}
\usepackage{dcolumn}
\usepackage{epsfig}
\usepackage{textcomp}
\usepackage{multirow}
\usepackage{color}
\usepackage{adjustbox}
\modulolinenumbers[5]
\begin{document}
\title{Time series analysis of temporal networks}
\author{Sandipan Sikdar\inst{1} \, Niloy Ganguly\inst{1} \and Animesh Mukherjee\inst{1}
}                     
\offprints{}          
\institute{\inst{1}Indian Institute of Technology Kharagpur, Kharagpur, India}
\date{Received: date / Revised version: date}
%
\abstract{
A common but an important feature of all real-world
networks is that they are temporal in nature, i.e., the network structure changes over time.
Due to this dynamic nature, it becomes difficult to propose suitable growth models 
that can explain the various important characteristic properties of these networks. 
In fact, in many application oriented studies only knowing these properties is sufficient. For instance, if one wishes to launch a targeted attack on a network, 
this can be done even without the knowledge of the full network structure; rather an estimate of some of the properties is sufficient enough to launch the attack.
We, in this paper show that even if the network structure at a future time point is not available one can still manage to estimate its properties.
We propose a novel method to map a temporal network to a set of time series instances, 
analyze them and using a standard forecast model of time series, 
try to predict the properties of a temporal network at a later time instance. To our aim, we consider 
eight properties such as number of active nodes, average degree, clustering coefficient etc. and apply our prediction framework on them. 
We mainly focus on the temporal network of human face-to-face contacts and observe that 
it represents a 
stochastic process with memory that can be modeled as Auto-Regressive-Integrated-Moving-Average (ARIMA). 
We use 
cross validation techniques to find the percentage accuracy of our predictions. 
An important observation is that the frequency domain properties of the time series obtained from spectrogram analysis could be used to refine the 
prediction framework by identifying beforehand the cases where the error in prediction is likely to be 
 high. This leads to an improvement of {\bf 7.96\%} (for error level $\leq 20\%$) in prediction accuracy on an average across all datsets.
As an application we show how such prediction scheme can be used to launch targeted attacks on temporal networks.
} 
\maketitle

\section{Introduction}
\label{introduction}
\if{0}
Networks have been extensively used for the past many years for analyzing biological, social and various other complex systems observed in nature.
Several growth models ~\cite{albert2002statistical,newman2000models} have been proposed and several special properties of these networks like small-world, scale-free have 
been identified. However, the underlying assumption in most of these works is that the networks 
 being analyzed are static. Only 
\fi
Recently the research community is reaching a consensus that many real 
networks have nodes and edges entering or leaving the system dynamically, thus introducing the dimension of time. This special class of networks are often called 
temporal networks ~\cite{Holme2012} or time-varying networks. Initial studies on temporal networks have been performed by aggregating the nodes and edges over all time 
steps and then analyzing the behavior of the aggregated network. This strategy however hides the time ordering of the nodes and the edges 
which may have a significant role in the understanding of 
the true nature of such temporal networks.
Researchers have subsequently come up with growth models and have proposed several metrics. Tang et al. have showed in ~\cite{TSMML10:smallworld} the presence of correlation and small world 
behavior in temporal networks. In ~\cite{stehle2010dynamical} the temporal network of human communication has been identified to be bursty in nature.
Recently, new network applications have cropped up where an estimate of the network properties are helpful even though the network structure is itself unavailable. 
For instance, in order to launch a targeted attack on the network one might not require the full knowledge of the network structure. Instead, an approximate estimate of some 
of the properties might be useful in finding the order in which the nodes and the edges may be removed. 

Our contributions in this paper are fourfold.
We propose a simple strategy to {\bf represent a temporal network as time series}. Essentially, we consider a temporal network as a set of static snapshots collected 
at consecutive time intervals and represent each of them in terms of the properties of the network. In specific, we consider eight properties namely number of 
active nodes, average degree, clustering coefficient, number of active edges, betweenness centrality, closeness 
centrality, modularity and edge-emergence ~\cite{sur2014attack}.

We then use the known analytical tools for time series predictions to {\bf predict the network properties at a future time instance.}  
Note that the time series framework can be particularly effective as it is impossible to 
define a unified network evolution/growth model for temporal networks simply because the  
rules of temporality are varied across systems. Hence the feasible alternative could be to learn 
the evolution pattern (which we do through time-series analysis) and then predict the later time steps.

Due to various irregularities in the time series, predictions at certain points are erroneous. Therefore we 
further {\bf refine our prediction framework using spectrogram analysis} by identifying beforehand the cases where the 
prediction error is high i.e., unsuitable for prediction. In fact we observe that the accuracy of the framework is enhanced further by {\bf 7.96\%} (for error level $\leq 20\%$) on an average across all datasets 
if we remove 
the cases which are deemed unsuitable for prediction by spectrogram analysis.

As an application we also {\bf propose a strategy to launch targeted attacks based on our prediction framework} and show that this scheme beats the state-of-the-art ranking method 
used for such attacks. We believe that our framework could be used in designing ranking schemes for nodes in temporal networks at a future time step  
albeit the network structure at that time step itself is unknown.


We perform our experiments on five different human face-to-face communication networks 
and observe that the above properties could be segregated based on time domain and frequency domain (spectrogram) characteristics. 
In general this method allows us to make predictions with very low errors.   
 Importantly, the frequency domain analysis also nicely separates out those properties that can be predicted with low errors from those for which it is not 
 possible.
Rest of the paper is organized as follows. In Section ~\ref{relatedworks} we present a brief review of the literature. 
Section ~\ref{mapping} describes the framework for mapping temporal networks to time series. 
Section ~\ref{dataset} provides a brief description of the datasets we use for 
our experiments.
  In section ~\ref{properties} we perform a detailed time domain and frequency domain analysis of the time series. 
Section ~\ref{prediction} outlines the description of our prediction framework. In Section ~\ref{result} we provide the detailed results of our prediction framework on the 
human face-to-face communication networks. We also show how the prediction scheme could be enhanced using the spectrogram analysis. We further 
propose an attack strategy based on the prediction scheme and show that it beats the state-of-the-art methods (section ~\ref{attack}). 
We conclude in section ~\ref{conclusion} by summarizing our main contributions and pointing to certain future directions.
\section{Related works}
\label{relatedworks}
Most of the initial works attempted to study temporal networks by aggregating the network across all times and then analyzing this aggregated network.
However, it was found that
time ordering is an important issue and destroying this ordering information severely affects the understanding of the true nature of the network. 
Several ways have been therefore devised to represent temporal networks. Basu et al. 
have shown in ~\cite{journals/corr/abs-1012-0260} a way of representing temporal networks as a time series of static graph snapshots and have proposed a stochastic model for generating 
temporal networks. Perra et al. in ~\cite{perra2012activity} have provided an activity driven modeling of temporal networks. They define activity potential which is a time 
invariant function characterizing the agents' interactions and propose a formal model of temporal networks based on this idea. Random walks have been introduced in 
temporal networks ~\cite{starnini2012random} to single out the role of the different properties of the empirical networks. It has also been shown that random walk 
exploration is slower on temporal networks than it is on the aggregate projected network. Several other modeling frameworks for temporal networks have also 
been proposed (see ~\cite{vespignani2011modelling,hill2010dynamic,hanneke2010discrete} for references). Apart from the attempts to generate temporal networks several other properties of these networks have also been investigated.
Temporal networks of human communication has been observed to be bursty in nature in ~\cite{barabasi2005origin}. Dynamics of human face-to-face interactions have been studied in ~\cite{zhao2011social}.
~\cite{tang2009temporal} shows how the presence of burstiness affects the dynamics of diffusion process. Further different metrics to study the properties of temporal networks 
have also been proposed in ~\cite{pan2011path}.  

On the other hand, time series have found a lot of applications in economic 
analysis and financial forecasting ~\cite{hamilton1989new}. It has also been applied in tweet analysis ~\cite{o2010tweets}. 
However, temporal networks have not 
been studied in details so far as a time series problem except for preliminary attempts in ~\cite{scherrer2008description,hempel2011inner}.
In ~\cite{scherrer2008description} the authors analyze temporal networks as time series but to the best of our knowledge this is the first 
work which leverages the time series forecasting tools to predict the network properties at a future time instant. 
Frequency domain analysis on the time series representing 
temporal network and its implication also remain unexplored to the best of our knowledge.
 Therefore we propose to leverage in this paper the standard 
techniques of time series analysis (both in time and frequency domain) to understand the dynamical properties of temporal networks.
Our prime contribution is to develop a unified framework to analyze and predict different properties of temporal networks based on time-series modeling.

\section{Mapping Temporal network to time series}
\label{mapping}

\begin{figure}
 \begin{center}
 \includegraphics*[width=0.7\columnwidth, angle=0]{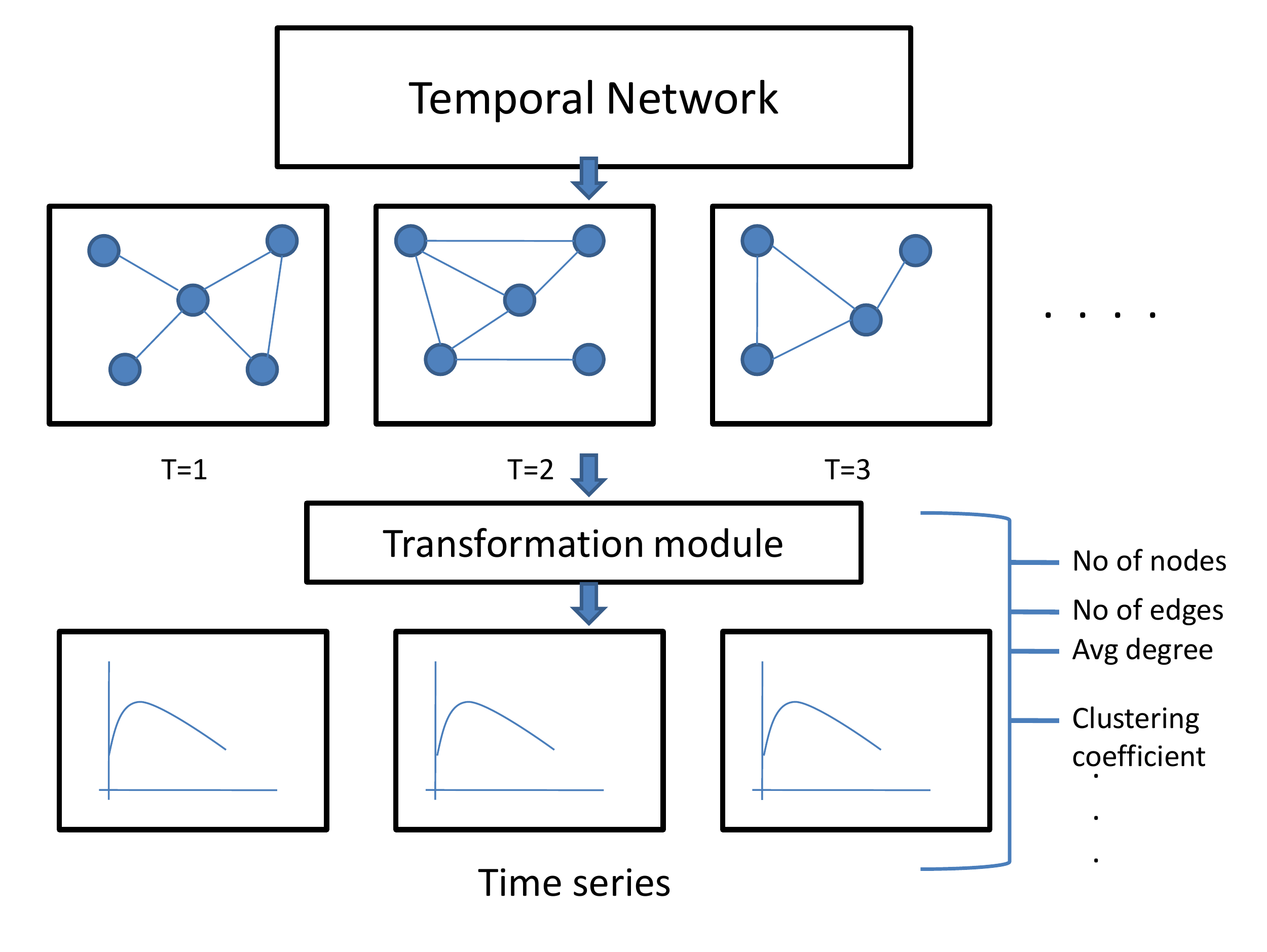}
 \caption{\label{fig1}Converting Temporal network to Time series}
 \end{center}
 \end{figure}

 We consider a temporal network as a set of static snapshots collected at consecutive time intervals. 
Each snapshot is then represented in terms of eight (mainly structural) properties of the underlying graph. 
 Consequently we obtain a set of points ordered in time or equivalently a discrete time series (see figure ~\ref{fig1}). 
The eight properties we use to represent the temporal network as a time series are:
 
 \begin{itemize} 
  \item (1). {\bf  Number of active nodes}: This is the count of the number of nodes in the system at a given time step. We consider active nodes to be those which have non-zero degree in a time step. We represent the number of active nodes in the system at time step {$t$} by {$N_{t}$}.
  In a similar way we define (2). {\bf number of active edges} and (3). {\bf average degree} and represent the values of these properties at time $t$ by $E_t$ and $Avg\_deg_{t}$ respectively.

\begin{figure}[h]
 \begin{center}
 
 \includegraphics[width=0.7\columnwidth, angle=0]{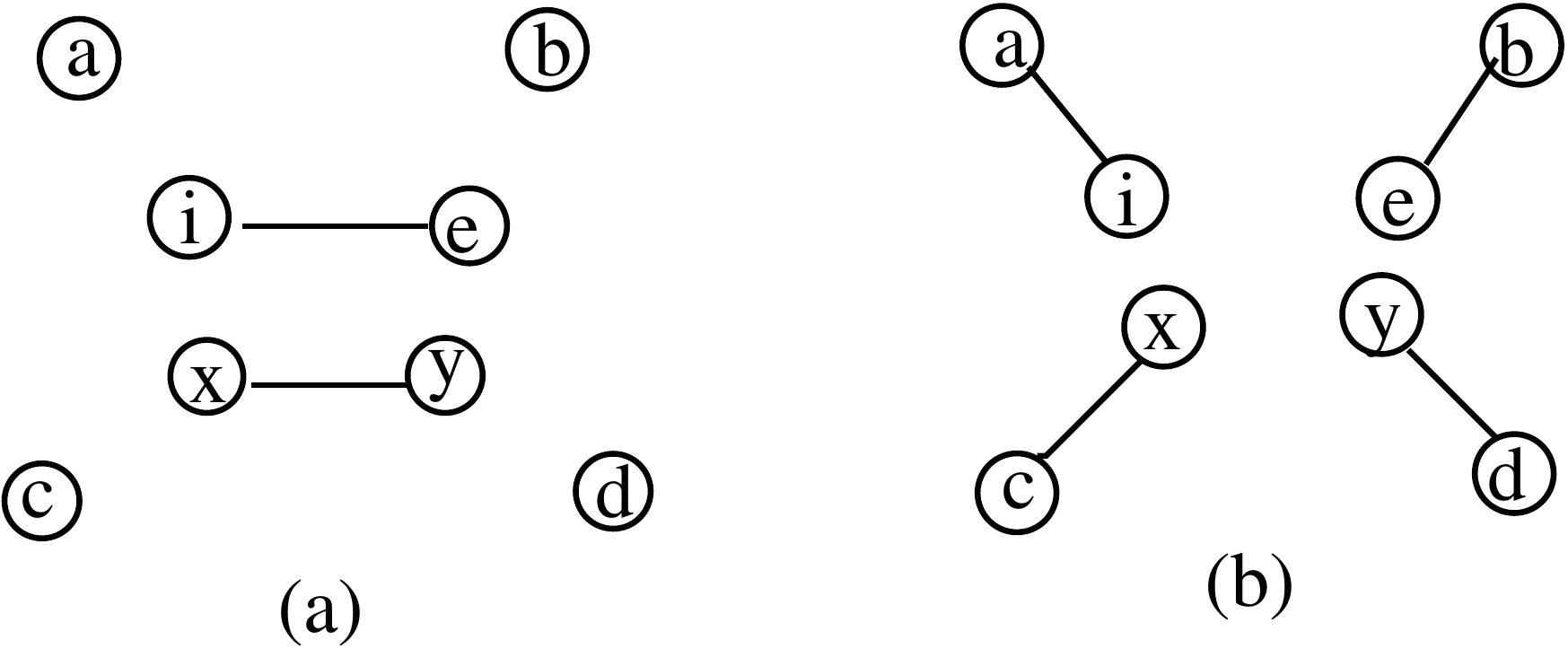}
 \caption{\label{fig3}(a) and (b) denote the status of the network at time $t$ and $t+1$ respectively. For the edge $(i,e)$ in $t$ the corresponding edges
 emanating from $i$ and $e$ are $(i,a)$ and $(e,b)$. For the edge $(x,y)$ they are $(x,c)$ and $(y,d)$. So the $Edge\_emer_{t}=\frac{2+2}{2}=\frac{4}{2}$}
 
%

 \end{center}
 \end{figure}

\item (4). {\bf  Edge emergence}: Edge emergence ~\cite{sur2014attack} is a measure that estimates structural similarity. For measuring the 
  edge emergence at time $t$ we consider each edge of the network at time $t$ and for each of its two endpoints we calculate the number of edges 
  emerging in the next time step $t+1$. We represent edge-emergence at time $t$ by $Edge\_emg_{t}$. If $E_t$ denotes the set of edges present in the network at time $t$ 
  and $A_{t+1}$ denotes the set of edges at time $t+1$ which are adjacent to $E_t$ then $Edge\_emg_{t}=\frac{|A_{t+1}|}{|E_{t}|}$. 
Figure~\ref{fig3} shows how we calculate this measure for a temporal network at any time instance.

  \item (5). {\bf  Modularity}: We decompose each snapshot into communities using the technique specified in ~\cite{blondel2008fast}
  and measure the goodness of this division using modularity ~\cite{newman2006modularity}. We represent modularity 
  of the system at a given time step {$t$} by {$Mod_{t}$}. 
  
  
 \end{itemize}
 We also consider (6). {\bf  betweenness centrality}, (7). {\bf closeness centrality} and (8). {\bf  clustering coefficient} of the graph (values computed for each node and then summed over all nodes) 
and their values at time step $t$ are represented by 
  {$Bet\_cen_{t}$}, {$Clos\_cen_{t}$} and {$Clus\_coeff_{t}$} respectively.

\section{Description of the dataset}
\label{dataset}
We perform our experiments on five human face-to-face network datasets: INFOCOM 2006 dataset~\cite{cambridge-haggle-imote-infocom2006-2009-05-29}, SIGCOMM 2009 dataset~\cite{thlab-sigcomm2009-mobiclique-proximity-2012-07-15}\footnote{http://crawdad.org/}, 
High school datasets (2011, 2012)~\cite{fournet2014contact} and Hospital dataset~\cite{vanhems2013estimating}\footnote{http://www.sociopatterns.org/}.

\begin{itemize}
\item {\bf INFOCOM 2006:}
This is a human face-to-face communication network and was collected at the IEEE INFOCOM 2006 conference at Barcelona.
78 researchers and students participated in the experiment. They were equipped with imotes and apart from them 20 stationary imotes were deployed as 
location anchors. The stationary imotes had more powerful battery and had a radio range of about 100 meters. The dynamic imotes had a radio range of 30 meters.
If two imotes came in each others' range and stayed for at least 20 seconds then an edge was recorded between the two imotes.
The edges were recorded at every 20 seconds. 
Therefore, this is the lowest resolution at which the experiments can be potentially conducted. However, we observe that at this resolution the network is extremely sparse 
which makes it difficult to conduct meaningful data comparison and prediction. We observe experimentally that the lowest interval that allows for appropriate comparison 
and prediction is $5$ minutes and therefore we set this value as our resolution for all further analysis.
 \item {\bf SIGCOMM 2009:}
 This is also a human face-to-face communication network and was collected at the SIGCOMM 2009 conference at Barcelona, Spain. The dataset contains data collected by an opportunistic mobile social 
application, MobiClique. The application was used by 76 persons during SIGCOMM 
2009 conference in Barcelona, Spain. The trace records all the nearby Bluetooth devices reported by the periodic 
Bluetooth device discoveries.
Each device performed a periodic Bluetooth device discovery every 120+/-10.24 seconds for nearby Bluetooth devices. A link was added with a device 
on discovering it. 
We remove the contacts with external Bluetooth 
devices and a network snapshot is an aggregate of data obtained for 5 minutes. 

\item {\bf High school datasets:}
These are two datasets containing the temporal network of contacts between students in a high school in Marseilles taken during December 2011 and November 2012 respectively. 
Contacts were recorded at intervals of 20 seconds. We consider a network snapshot as an aggregate of data obtained for 5 minutes. 

\item {\bf Hospital dataset:}
This dataset consists of the temporal network of contacts between patients and health care workers in a hospital ward in Lyon, france. Data was collected at every 20 second intervals.
 Due to sparseness of the network of 20 seconds, we consider each network snapshot as an aggregated network of 5 minutes.

 In table 1 we provide the details of the datasets.  
%
 \end{itemize}
 
\begin{table*}
\centering
\begin{adjustbox}{max width=\textwidth}
\begin{tabular}{|c|c|c|c|c|c|}
\hline
Dataset         & \# unique nodes & \# unique edges & edge type  & \begin{tabular}[c]{@{}l@{}}Time span of \\ the dataset\end{tabular}     & \begin{tabular}[c]{@{}l@{}}Time steps \\ for  prediction\end{tabular} \\ \hline
INFOCOM 2006    & 98              & 4414            & undirected &  1120                                                                   & 200 - 800                                                             \\ \hline
SIGCOMM 2009    & 76              & 2082            & do         &  1068                                                                   & 300 - 900                                                             \\ \hline
Highschool 2011 & 126             & 5758            & do         &  1215                                                                   & 200 - 900                                                           \\ \hline
Highschool 2012 & 180             & 8384            & do         &  1512                                                                   & 200 - 1000                                                            \\ \hline
Hospital        & 75              & 5704            & do         &  1158                                                                   & 100 - 900                                                             \\ \hline
\end{tabular}
\end{adjustbox}
\label{tab_data}
\caption{Properties of the dataset used.}
\end{table*}

 \noindent
\section{Analysis of time series}
\label{properties}

In this section we present the plots of the time series and analyze their properties based on both time domain and frequency domain characteristics. 
\begin{figure*}[!ht]
  \centering
  \includegraphics*[width=0.9\textwidth,angle=0]{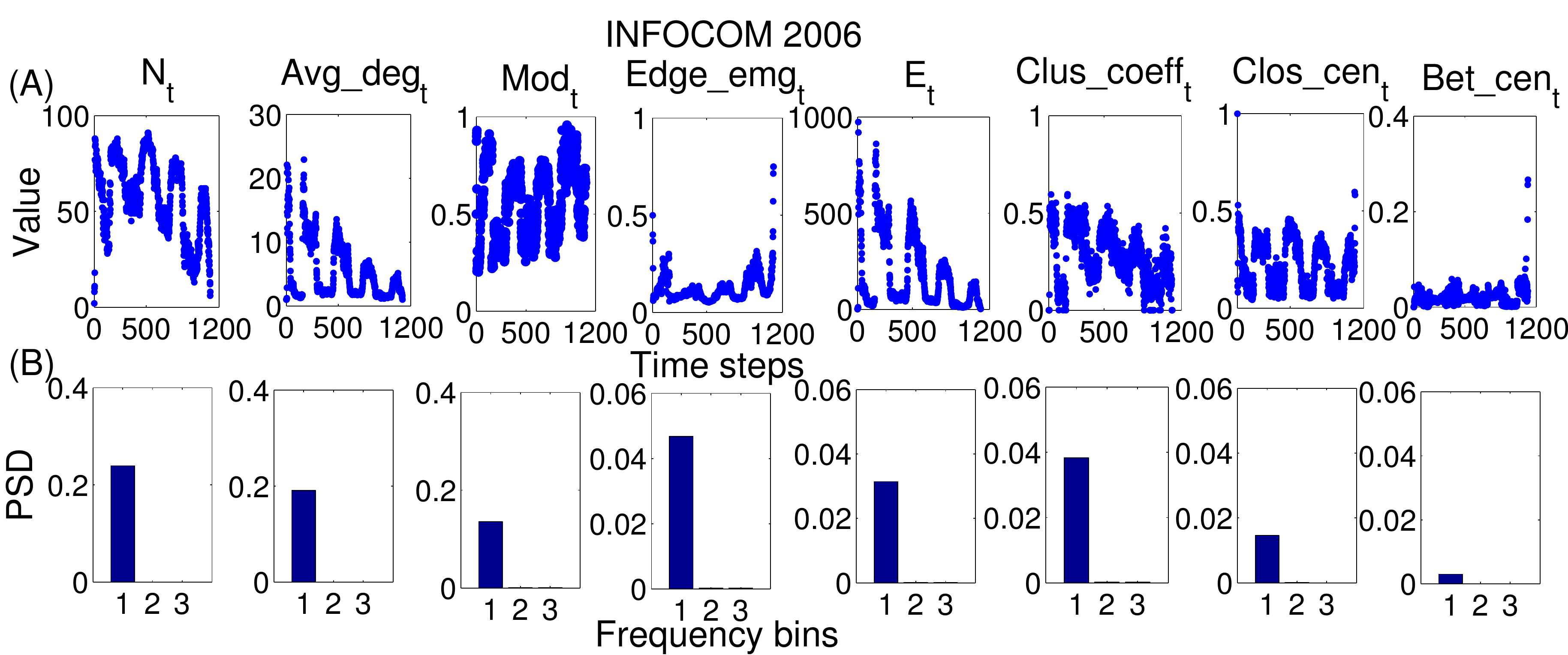}
  
  

  \centering
  \includegraphics*[width=0.9\textwidth,angle=0]{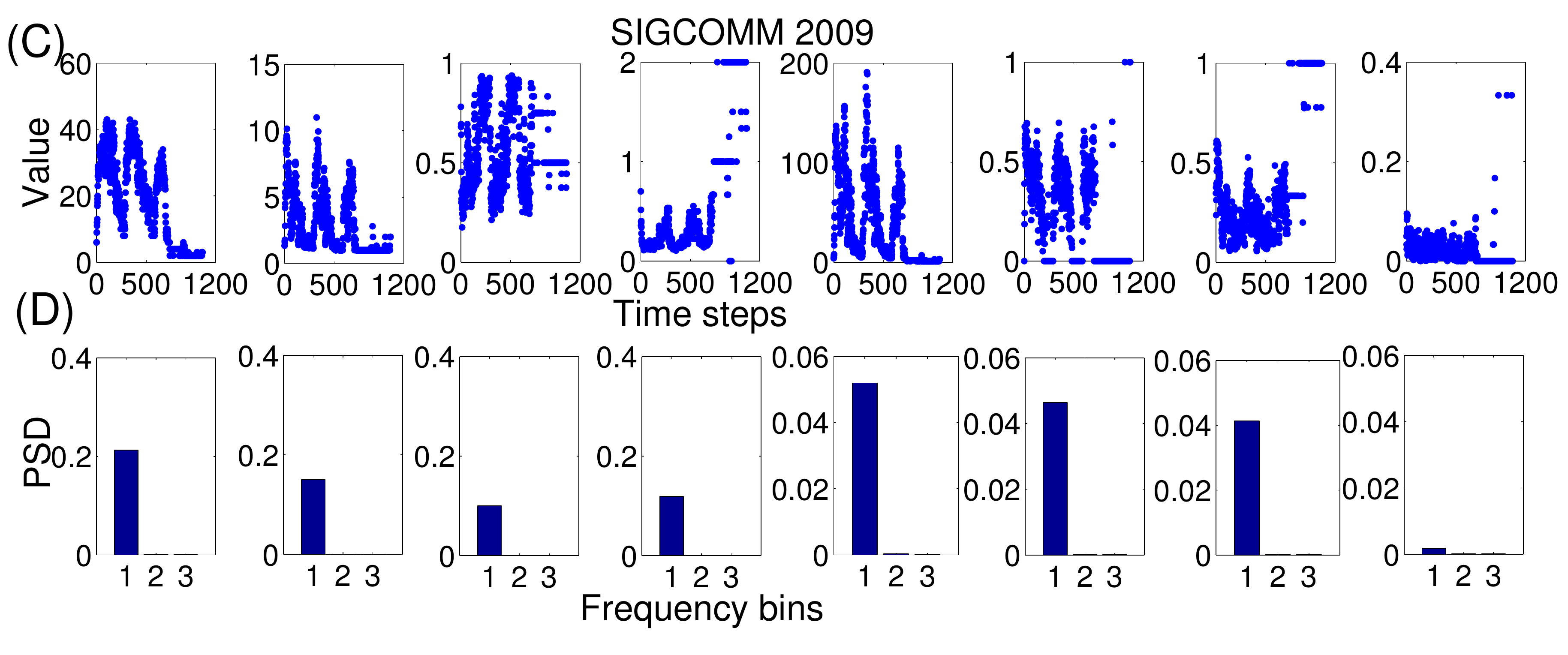}
  
  \centering
  \includegraphics*[width=0.95\textwidth,angle=0]{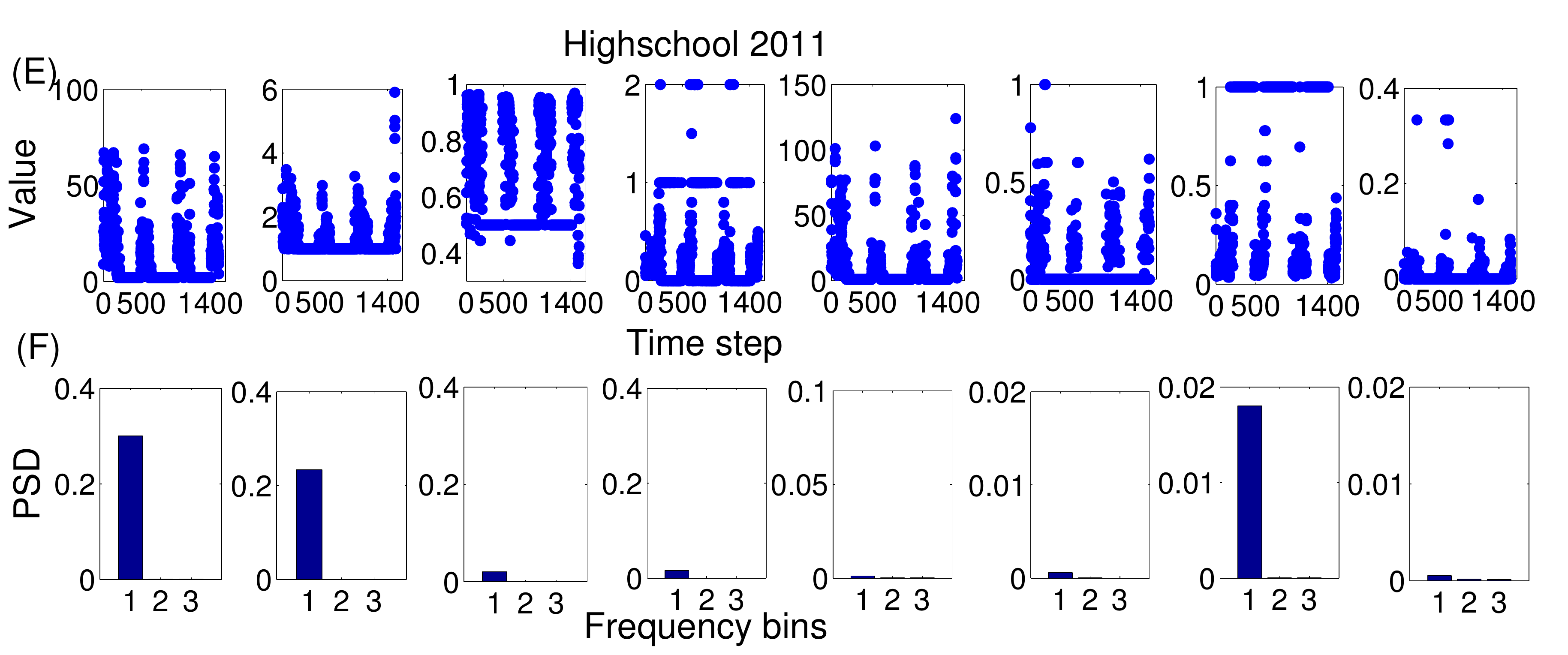}
 
   \caption{\label{fig_all_dataset} (A), (C) and (E) represent the time series plots for INFOCOM 2006, SIGCOMM 2009 and High-school 2011 respectively. (B), (D) and (F) represent the power spectral density (PSD) corresponding to the frequency bins for INFOCOM 2006, SIGCOMM 2009 and High-school 2011 dataset respectively. Bins 1, 2 and 3 corresponds to frequencies $<$5, 5-15 and $>$15(Hz) respectively.}
 \end{figure*}
  \begin{figure*}[!ht]
  
    \centering
  \includegraphics*[width=0.95\textwidth,angle=0]{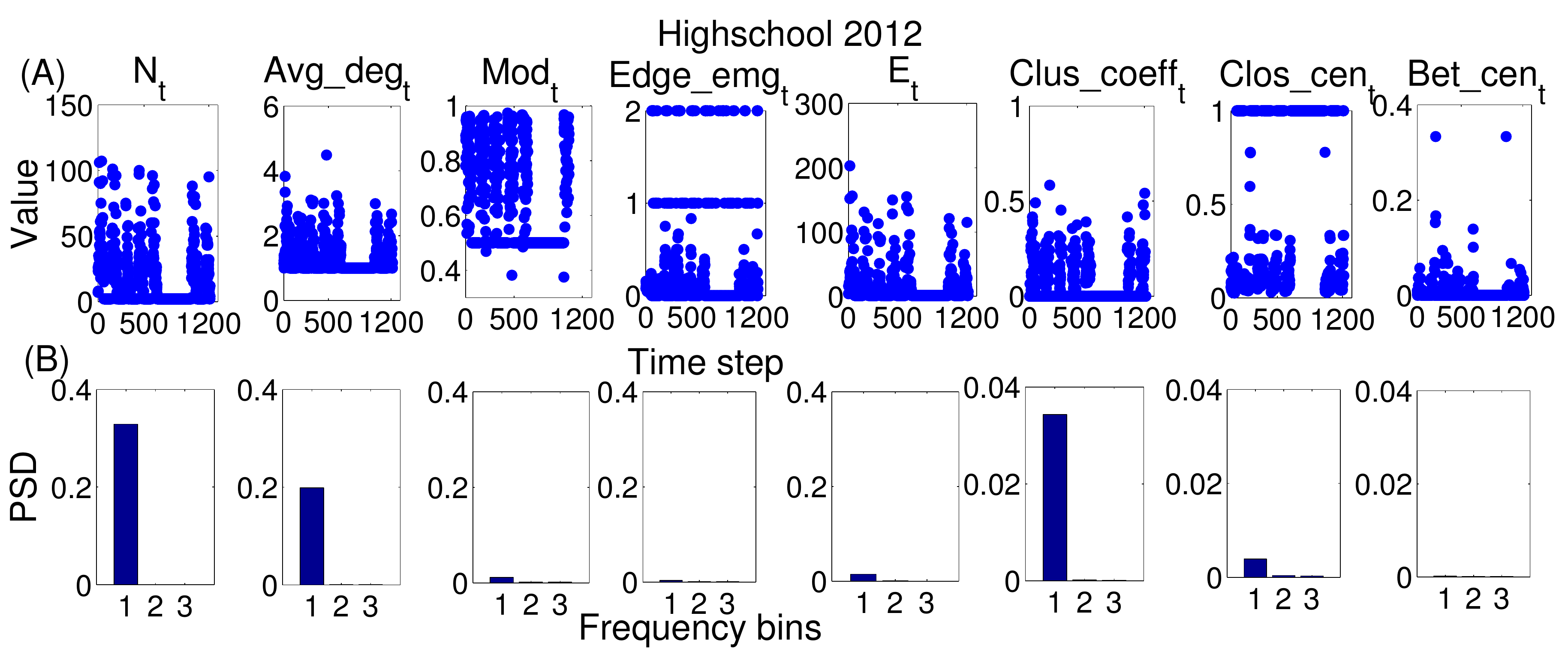}
  
  \centering
  \includegraphics*[width=0.95\textwidth,angle=0]{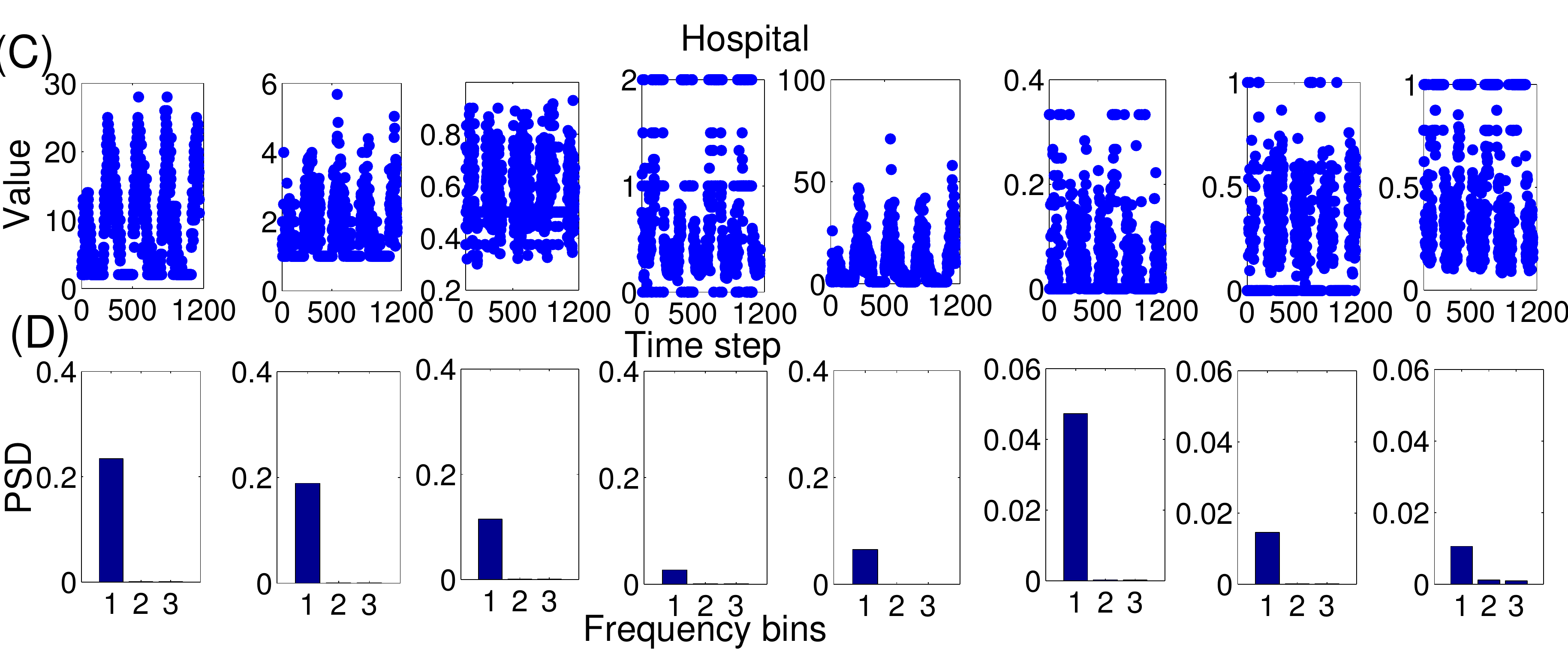}
  
  \caption{\label{fig_all_dataset_1} (A) and (C) represent the time series plots for High-school 2012, Hospital respectively. (B) and (D) represent the power spectral density (PSD) corresponding to the frequency bins for High-school 2012 and Hospital dataset respectively. Bins 1, 2 and 3 corresponds to frequencies $<$5, 5-15 and $>$15(Hz) respectively.}
  
  \end{figure*}

\subsection{Time domain characteristics}

For the time domain analysis of the properties we look into the time series plots for the datasets represented in figures ~\ref{fig_all_dataset}(A), (C), (E) and 
~\ref{fig_all_dataset_1}(A), (C).
From the time series plots we observe the presence of periodicity in almost all the datasets. A stretch of high values is followed by a stretch of low values 
and so on. However, they are of varying lengths.  
This indicates the presence of correlation in case of human face-to-face communication network.
We quantify this structural correlation later in this paper. 
We also check whether these time series are stationary. 
On performing KPSS (Kwiatkowski–Phillips–Schmidt–Shin) ~\cite{kwiatkowski1992testing} and 
 ADF (Augmented Dickey Fuller) test ~\cite{dickey1979distribution} on the data we conclude that the data is non-stationary. 
 Overall, the presence of correlation in case of human face-to-face network indicates that it is a stochastic process with memory i.e., the contacts a node 
 makes in the current time step is influenced by its contact history.

\subsection{Frequency domain analysis}
\label{spectrogram}

In this section, we perform the frequency domain analysis of 
the time series extracted from the temporal network
by conducting
  a spectrogram analysis of the data. Spectrogram analysis is a short time fourier transform 
where we divide the whole time series into several equal sized windows and apply discrete fourier transform on this widowed data.
The main advantages of using spectrogram analysis are
 (a). we do not lose the time information,
 (b). we are able to obtain a view of the local frequency spectrum.
Also note that the spectrogram analysis allows us to identify as well as quantify the fluctuations in the data which is difficult to identify 
from the corresponding time series. 
A high concentration of low frequency components would indicate lower fluctuations in the data; in contrast no such concentration of low frequency components 
would indicate higher fluctuations and irregularities in the data.

We construct the spectrogram and segregate the power spectral density (PSD measured in Watts/Hz) based on the frequency into three bins. In bin 1 we calculate the mean PSD 
corresponding to the frequencies $<$ 5 Hz, in bin 2 we calculate the PSD corresponding to frequencies between 5 and 15 Hz and bin 3 consists of the 
mean PSD value corresponding to frequencies $>$ 15 Hz. We call them LPSD, MPSD and HPSD respectively.
So a higher value of mean PSD corresponding to bin 1 (LPSD) would indicate lower fluctuations in data. 
In figures ~\ref{fig_all_dataset}(B), (D), (E) and ~\ref{fig_all_dataset_1}(B), (D) we plot the PSD corresponding to the three bins across all the properties for all the datasets. We 
observe that the lower frequencies dominate to a higher extent in case of the properties like number of active nodes, number of active edges, modularity 
but to a much lower extent in case of betweenness centrality, closeness centrality and clustering coefficient. We show later in this paper that 
the prediction accuracy of a property can be enhanced through spectrogram analysis.

\section{Prediction framework}
 \label{prediction}

\begin{figure}
 \begin{center}
 \includegraphics[width=0.45\columnwidth, angle=0]{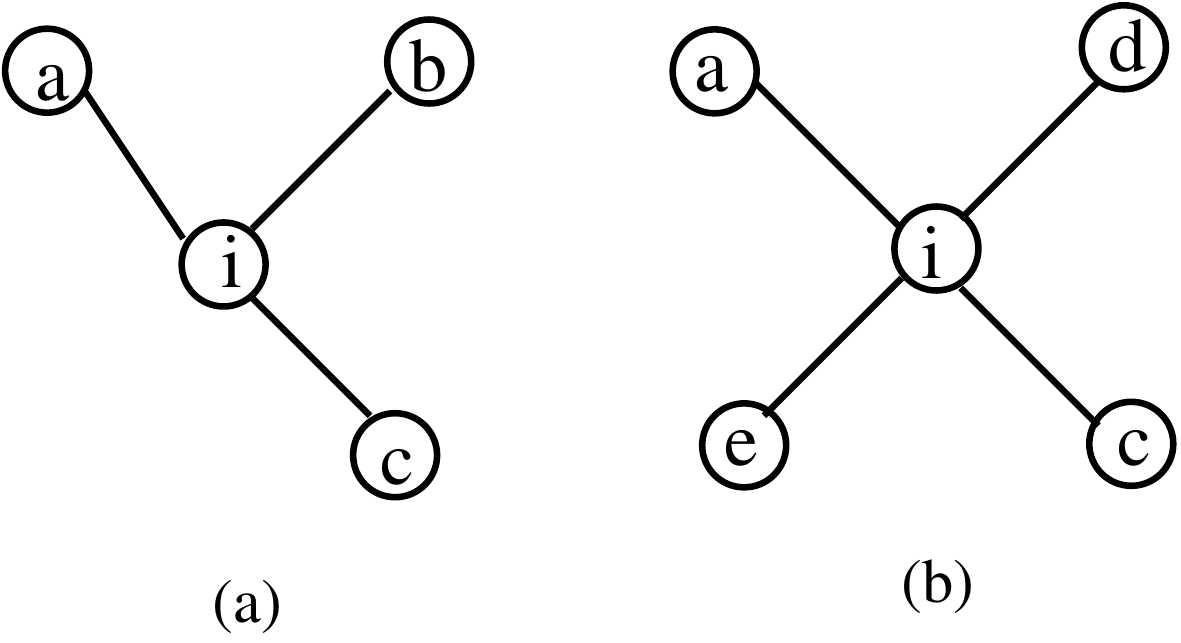}
 \caption{\label{fig2}(a) and (b) denote the status of node $i$ at time $t$ and $t+k$ respectively. $NBR(i)_{t}=\{a,b,c\}$ and $NBR(i)_{t+k}=\{a,d,c,e\}$ 
 $Correlation(i)_{k}$={\large$\frac{NBR(i)_{t}\bigcap NBR(i)_{t+k}}{NBR(i)_{t}\bigcup NBR(i)_{t+k}}$}={\large$\frac{2}{5}$} where $NBR(i)_{t}\rightarrow$ the set of neighbors of $i$ at time $t$}
  \end{center}
 \end{figure}
  
 \begin{figure*}
 \centering
   \includegraphics*[width=0.75\textwidth,angle=0]{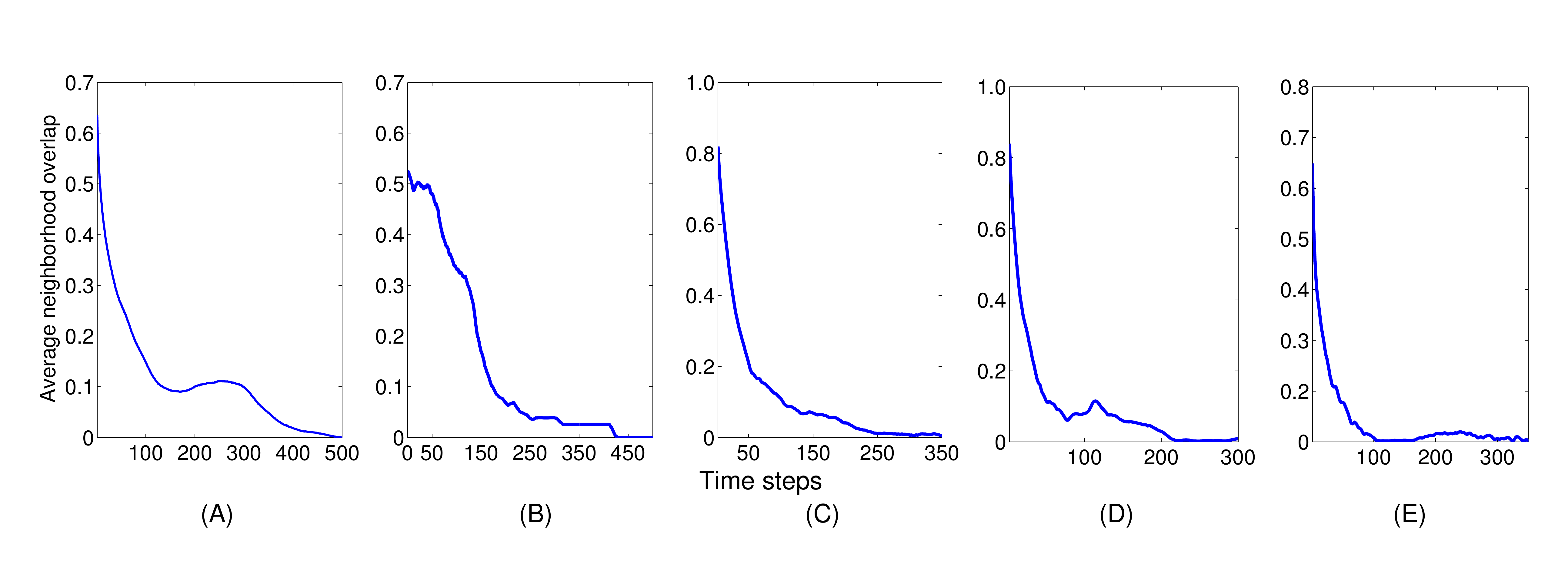}
 \caption{\label{aging}The average neighborhood-overlap value at different lags for (A)INFOCOM 2006, (B)SIGCOMM 2009, (C)Highschool 2012, (D)Highschool 2011 and (E)Hospital datasets.}
 \end{figure*}

In this section, we employ the time series to forecast the different structural properties of the temporal networks.
Elementary models of time series forecasting could be categorized into Auto-regressive(AR) and Moving average(MA) models ~\cite{chatfield2013analysis}. In case of 
an auto-regressive model of order $p$, AR($p$), the value of the time series at time step $t$ is given as - 
\begin{center}
 $y_{t}=\alpha_{1}y_{t-1}+\cdots+\alpha_{p}y_{t-p}+e_t+c$
\end{center}
where $\alpha_{i}$s are parameters, $e_t$ is the white noise error term and $c$ is a constant.
Similarly, in case of Moving average model of order $q$, MA($q$), the value of the time series at time step $t$ is given as - 
\begin{center}
 $y_{t}=\beta_{1}e_{t-1}+\cdots+\beta_{q}e_{t-q}+\mu+e_t+c$
\end{center}
where $\beta_{i}$s are parameters, $e_t,e_{t-1},...$ are white noise error terms and $\mu$ is the expectation of $y_t$.
These two models could be combined into Auto-regressive-moving-average (ARMA($p$,$q$)) ~\cite{chatfield2013analysis} where the value of the time series at time step $t$ is given as
\begin{center}
 $y_{t}=\alpha_{1}y_{t-1}+\cdots+\alpha_{p}y_{t-p}+\beta_{1}e_{t-1}+\cdots+\beta_{q}e_{t-q}+e_t+c$
\end{center}

However, in our case the time series show evidences of non-stationarity and short term dependencies and these models are insufficient and hence we use ARIMA model ~\cite{box2011time} for forecasting. The initial differncing step in ARIMA model 
is used to reduce the non-stationarity.
On fitting an ARIMA($p$,$d$,$q$) model to a time series we obtain an auto-regressive 
equation of the form-

\begin{center}
 $y_{t}=\alpha_{1}y_{t-1}+\cdots+\alpha_{p}y_{t-p}+\beta_{1}e_{t-1}+\cdots+\beta_{q}e_{t-q}+c$
\end{center}

Hence we can take a time series corresponding to a network property and fit an ARIMA model to it. Thus, we obtain an auto-regressive equation for that series which 
can be used 
in forecasting. 
In order to predict a value at a future time point, we divide the data in smaller parts and perform our predictions on these smaller stretches.
In the next subsection we discuss how we perform this division.

\subsection{Selecting a window}

\begin{figure}
 \begin{center}
  \includegraphics[width=0.75\columnwidth, angle=0]{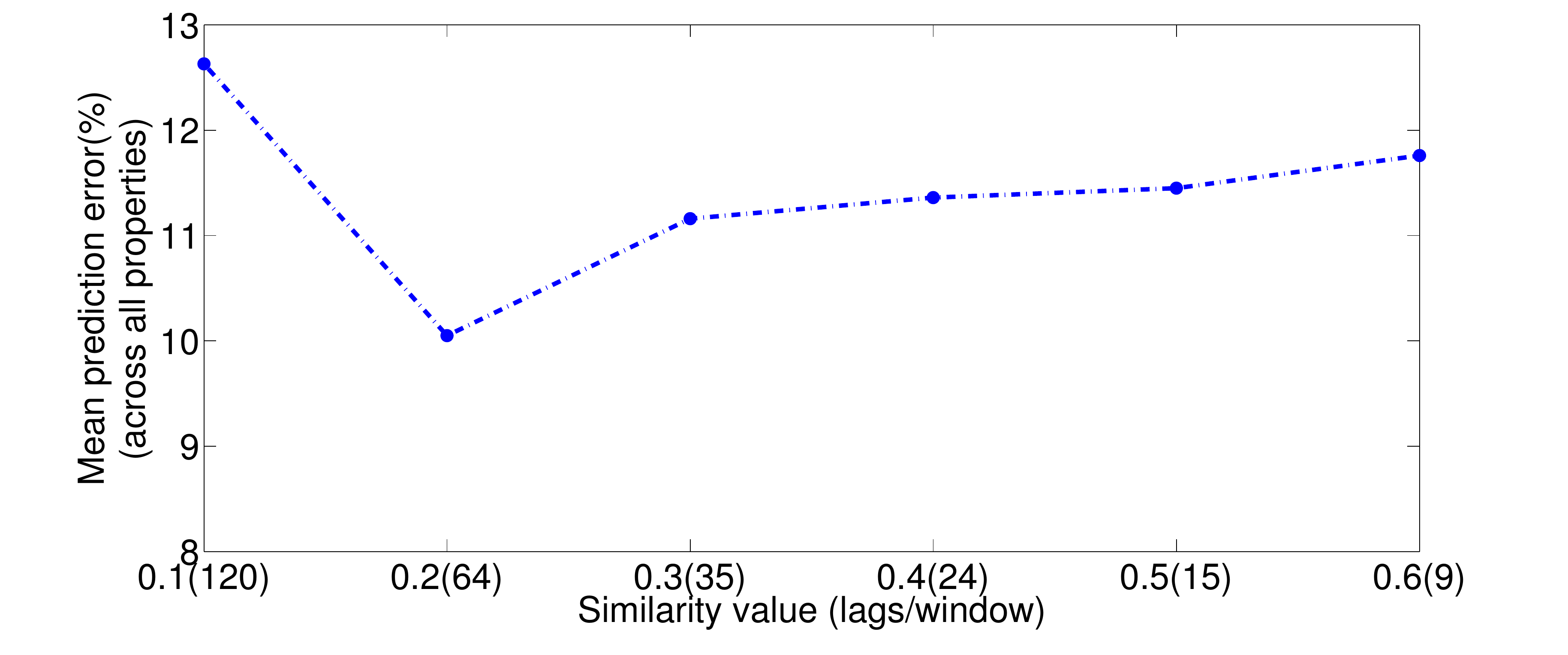}
  \caption{\label{fig_err} Mean prediction error (\%) across different properties for INFOCOM 2006 dataset for different similarity values. The lags corresponding to the similarity 
  value are also provided.}
  \end{center}
 \end{figure}

In order to identify the right length of a stretch (i.e., a window size) we need to identify how the network at any time point is influenced by the network at the previous time points.
The basic idea is that the time points to which this influence extends should all get included into a single window.
To quantify this influence we define a new metric called {\bf neighborhood-overlap} which measures the structural correlation between network snapshots at 
two time steps. We define the difference between these two time steps as the lag. To measure the neighborhood-overlap of the network snapshots at time 
$t$ and $t+k$, we calculate for each active node at time $t$ the overlap in its neighborhood between two time points. 
To measure this overlap we use the 
  standard Jaccard similarity as has been pointed out in ~\cite{TSMML10:smallworld}. Note that this is one of the most standard and interpretable ways to 
  measure structural similarity as has been identified in the literature with applications ranging from measuring keyword similarity ~\cite{niwattanakul2013using} to 
  similarity search in locality-sensitive-hashing (LSH) ~\cite{bawa2005lsh}. It has also been extensively used in link prediction ~\cite{lu2011link,lu2009similarity} 
  as well as community detection ~\cite{pan2010detecting}.
Figure~\ref{fig2} shows how we formulate this measure using the Jaccard similarity index. We represent the neighborhood overlap at lag $k$ as the mean value 
  across all the active nodes in time step $t$. 
  To measure the extent of similarity we measure neighborhood-overlap for each snapshot at different lags and take the average of them. This essentially 
  shows, given a time specific snapshot how the similarity changes as we increase the lag. Figures ~\ref{aging}(A) - (E) show how this similarity 
  changes with time as we increase lag for the five different datasets. 
  As we increase the lag the similarity  
 decreases almost exponentially and hence considering snapshots at larger lag where the similarity value is very low could introduce error in learning the auto-regressive equation. 
 Also for a higher similarity value the corresponding lag would increasingly introduce more error in the fit due to lesser number of data points on which the ARIMA model gets trained
 to learn the fit function (see figure ~\ref{fig_err}).
 In fact we observed that the error in prediction increases if we consider a lag too small (high similarity value) or too large (low similarity value) (see figure ~\ref{fig_err}). 
 Hence we consider the similarity value of $0.2$ as the threshold for calculating the lag. For our prediction framework the corresponding value of the lag acts as 
 the window for fitting the ARIMA model.
  
  Let the size of the window be $w$ and we want to predict the value of the time series at time $t$. To our aim we consider the time series of the previous $w$ 
  time steps consisting of the values between time steps $t-1-w$ to $t-1$ and fit the ARIMA model to it and obtain its value at time step $t$. 
  We repeat this procedure for forecasting at every value of $t$. Thus, the time step $t$ is the test point and the series of points $t-w-1$ to $t-1$ form the 
  training set. One can imagine this process as a sliding window of size $w$ which is used for learning the auto-regressive equation and the point that falls immediately 
  outside the window is the unknown that is to be predicted.
  
 \begin{figure*}
  \centering
  \includegraphics*[width=0.95\textwidth,angle=0]{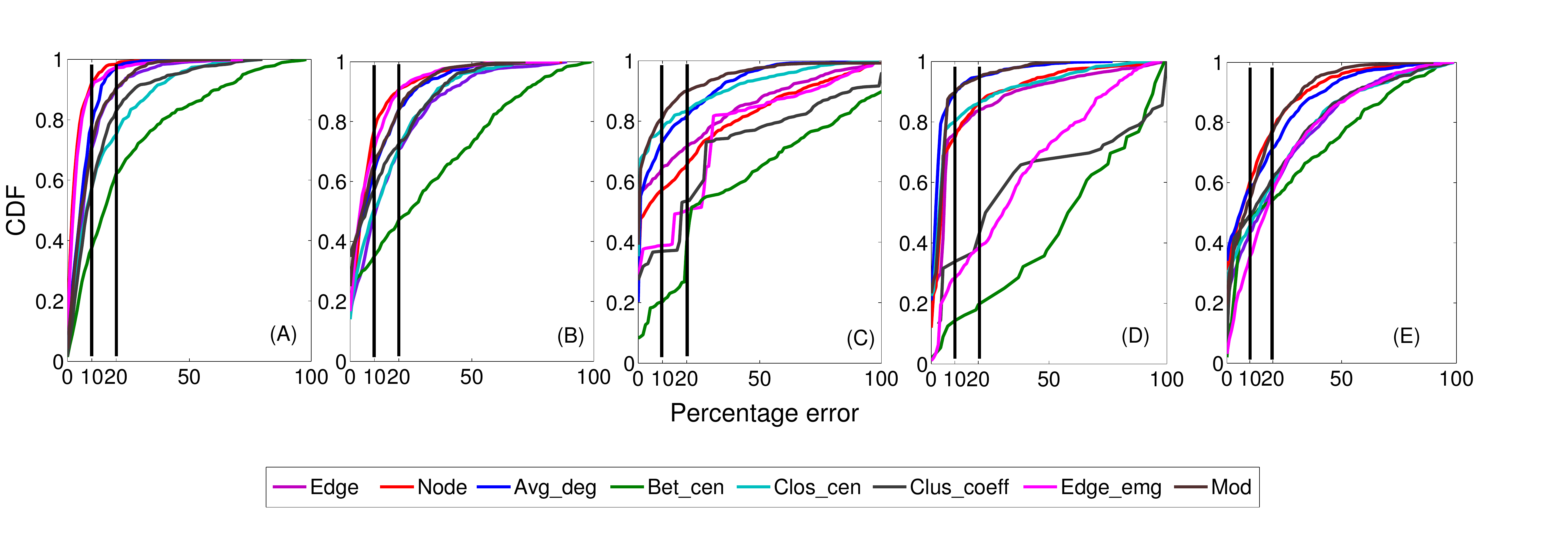}
 
 
 \caption{\label{fig8}  The percentage error distribution of all the properties (time series) for (A) INFOCOM 2006 dataset, (B) SIGCOMM 2009 dataset, (C) High-school 2011. (D) High-school 2012 and (E) Hospital. 
 X-axis represents percentage error and Y-axis represents 
 probability.}
\end{figure*}  

\section{Prediction results}
\label{result}
In this section, we provide detailed results of the our prediction framework on the datasets discussed earlier.
To determine the accuracy of our prediction strategy we use the cross validation technique.  
For each time step in this range we 
use our framework to obtain a prediction at that time step. Since we already know the original value, we can obtain a percentage error for the prediction.
Let $predict_{t}$ represent the prediction value at time $t$ and $original_{t}$ represent the original value. We obtain percentage error ($error_{t}$) using the formula:
\begin{center}
 {\large$error_{t}=\frac{|original_{t}-predict_{t}|}{original_{t}}*100$}
\end{center}

First we try to find the suitable window for predicting the value of a time series at a time step. For this we refer to figure ~\ref{aging} where 
we quantify structural correlation and show how the similarity value decreases with increasing lag.
We observe that the value of the structural correlation decreases as we increase the lag. For INFOCOM 2006 dataset (figure ~\ref{aging}(A)) the correlation drops to less than 
$0.2$ at lag around $70$.
Therefore we select a window of size 64. 
We could have selected any other value between 60 and 70, but we select 64 as it is in the power of 2 and it helps in the spectrogram 
analysis. Similarly we find the suitable window size to be around 128, 64, 64, 32 (closest power of 2) for the SIGCOMM 2009, Highschool 2011, Highschool 2012 and Hospital datasets respectively (refer to figure ~\ref{aging}). 

For the INFOCOM 2006 dataset we consider the time steps 200-800. 
Note that selection of these is just representative and one is free to take 
any time step given there is a window of appropriate length available. For SIGCOMM 2009, High School 2012, High School 2011 and Hospital datasets we 
consider our test time steps to be 300-900, 200-1000, 200-1000 and 100-900 respectively (refer to table 1). 

\begin{table*}
\centering
\begin{adjustbox}{max width=\textwidth}
\begin{tabular}{|c|c|c|c|c|c|}
\hline
                                                                  & \multicolumn{5}{c|}{Prediction error $\leq 20\%$}                                                                                                                                                                                                        \\ \hline
Datasets                                                          & \begin{tabular}[l]{@{}l@{}}INFOCOM \\ 2006\end{tabular} & \begin{tabular}[l]{@{}l@{}}SIGCOMM \\ 2009\end{tabular} & \begin{tabular}[c]{@{}l@{}}Highschool\\ 2012\end{tabular} & \begin{tabular}[c]{@{}l@{}}Highschool\\ 2011\end{tabular} & Hospital \\ \hline
\# Active nodes                                                   & {\bf 0.984}, (0.988)                                                  & {\bf 0.907}, (0.91)                                                   & 0.68, (0.765)                                                      & {\bf 0.861}, (0.882)                                                     & 0.782, (\underline{\it 0.859})    \\ \hline
Average degree                                                    & {\bf 0.975}, (0.968)                                                  & {\bf 0.84}, (0.834)                                                    & {\bf 0.816}, (0.81)                                                     & {\bf 0.91}, (0.908)                                                      & 0.714, (0.724)    \\ \hline
Modularity                                                        & {\bf 0.905}, (0.921)                                                  & {\bf 0.838}, (0.85)                                                   & {\bf 0.90}, (0.91)                                                      & {\bf 0.92}, (0.917)                                                      & 0.78, (0.812)     \\ \hline
Edge emergence                                                    & {\bf 0.971}, (0.983)                                                  & {\bf 0.906}, (0.91)                                                   & 0.56, (\underline{\it 0.71})                                                      & 0.42, (\underline{\it 0.512})                                                      & 0.57, (\underline{\it 0.652})     \\ \hline
\# Active edges                                                   & {\bf 0.901}, (0.91)                                                   & 0.71, (\underline{\it 0.81})                                                    & 0.72, (\underline{\it 0.78})                                                      & {\bf 0.836}, (0.86)                                                     & 0.734, (\underline{\it 0.796})    \\ \hline
\begin{tabular}[c]{@{}l@{}}Clustering \\ coefficient\end{tabular} & {\bf 0.829}, (0.858)                                                  & 0.725, (0.75)                                                   & 0.54, (\underline{\it 0.623})                                                      & 0.5, (\underline{\it 0.682})                                                       & 0.71, (\underline{\it 0.751})     \\ \hline
\begin{tabular}[c]{@{}l@{}}Closeness \\ centrality\end{tabular}   & 0.751, (\underline{\it 0.887})                                                  & 0.71, (\underline{\it 0.83})                                                    & {\bf 0.83}, (0.843)                                                      & {\bf 0.821}, (0.853)                                                     & 0.74, (\underline{\it 0.786})     \\ \hline
\begin{tabular}[c]{@{}l@{}}Betweenness\\ centrality\end{tabular}  & 0.621, (\underline{\it 0.818})                                                  & 0.472, (\underline{\it 0.61})                                                   & 0.51, (0.63)                                                      & 0.22, (\underline{\it 0.418})                                                      & 0.542, (\underline{\it 0.689})    \\ \hline
\hline
Average                                                           & 0.867, (\underline{\it 0.916})                                                  & 0.763, (\underline{\it 0.813})                                                   & 0.694, (\underline{\it 0.768})                                                     & 0.686, (\underline{\it 0.754})                                                     & 0.69, (\underline{\it 0.74})    \\ \hline
\end{tabular}
\end{adjustbox}
 \caption{\label{tab_res}Network property and the fraction of predictions with percentage error  $\leq20\%$ without (with) spectrogram analysis. 
 The cases where more than $80\%$ of the points have prediction error $\leq 20\%$ have been highlighted in bold font and the cases where on using spectrogram analysis 
 the improvement is more than $5\%$ have been underlined.}

\end{table*}
\begin{figure}
 \begin{center}
 \includegraphics[width=0.89\columnwidth, angle=0]{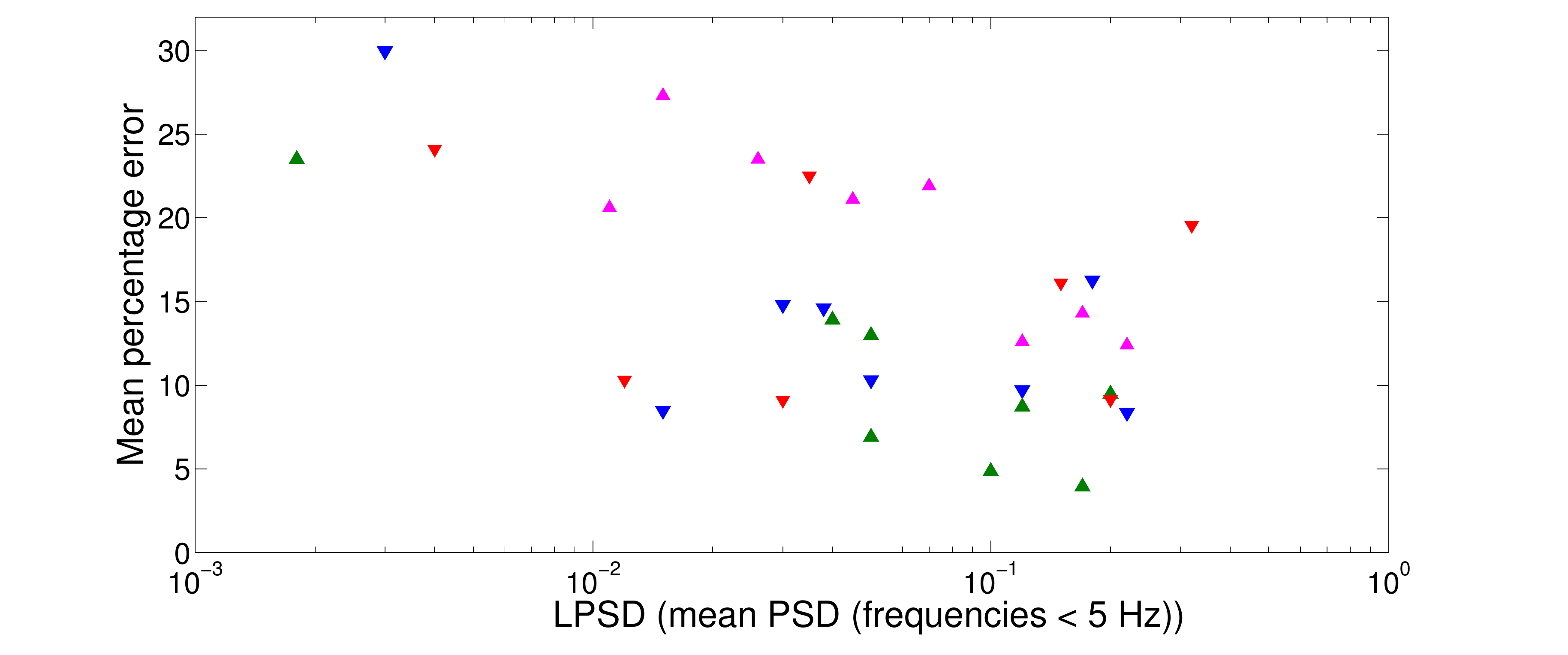}
 \caption{\label{k1}(A) LPSD versus mean percentage error for all the properties across all the datasets.}
  \end{center}
 \end{figure}

 \begin{figure}[!ht]
  \begin{center}
  \includegraphics[width=0.7\columnwidth, angle=0]{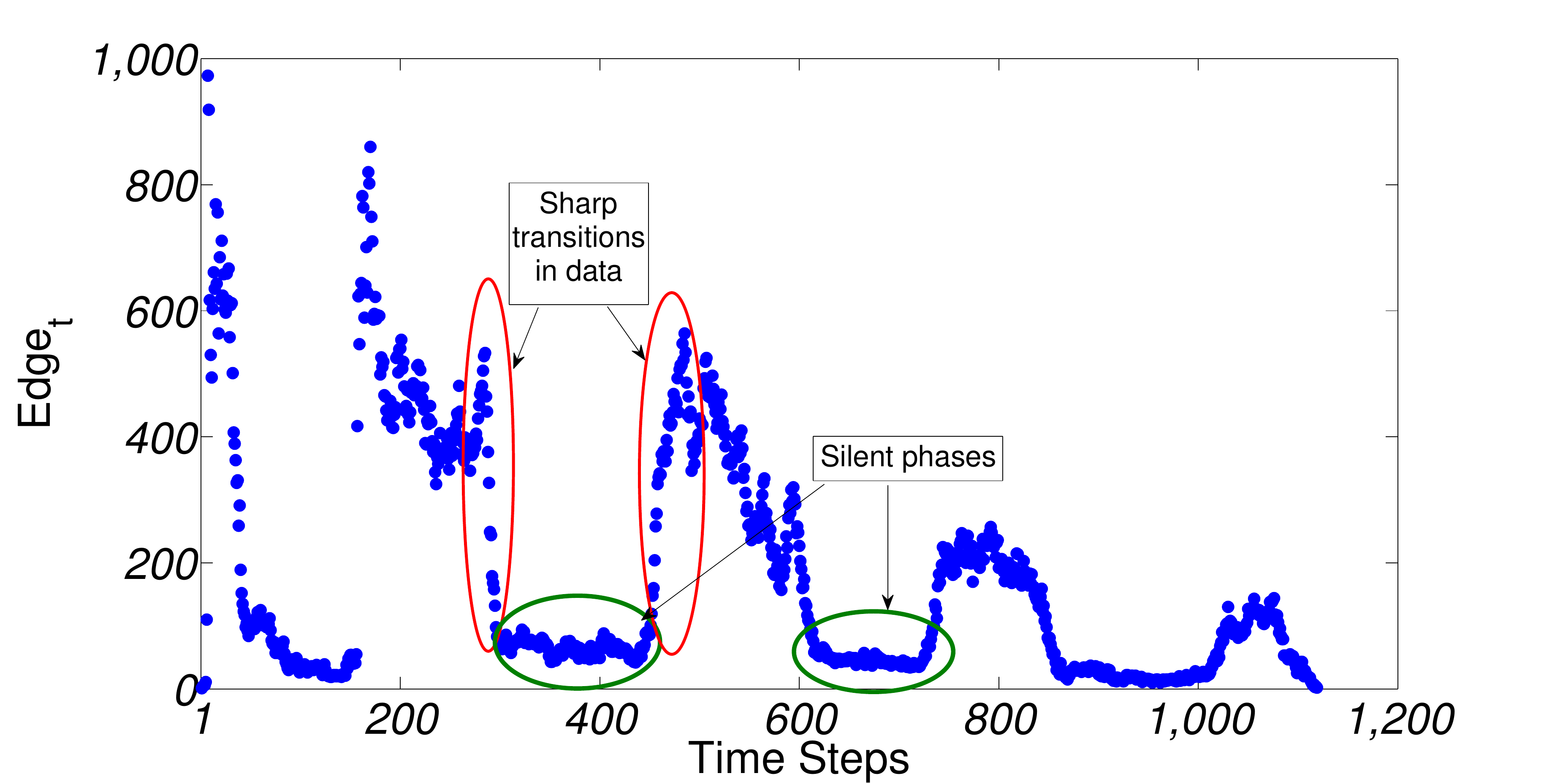}
  \caption{\label{fig9}The time series plot for number of active edges. The red and the green ellipses identify
  two transition and two silent phases respectively}
  \end{center}
 \end{figure}

 
 \begin{figure}[!ht]
  \begin{center}
  \includegraphics[width=0.95\columnwidth, angle=0]{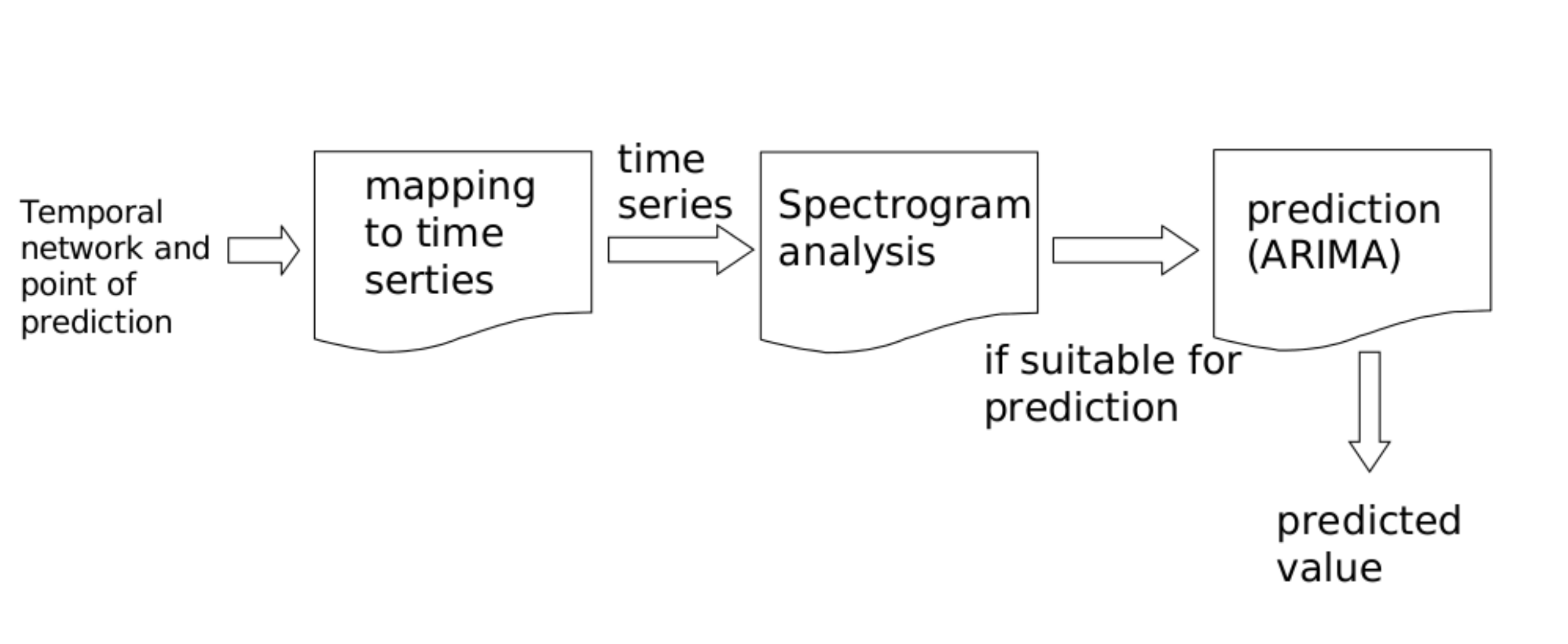}
  \caption{\label{fig13}The prediction framework.}
  \end{center}
 \end{figure}
 

To check how efficient our predictions are we plot the cumulative probability distribution of percentage error for all the datasets in figure ~\ref{fig8}. 
In table ~\ref{tab_res}, we compare the prediction results across different datasets and different metrics for cases where the prediction error $\leq20\%$. 
 Note that this error level is representative and ideally a table can be recovered for each such error level from figure ~\ref{fig8}. We make the 
 following observations from the results: 
\begin{itemize}
 \item  Our framework is able to predict the values for active nodes, average degree and modularity with high accuracy across all datasets. 
 \item For active edges, edge emergence, clustering coefficient and closeness centrality our framework is able to predict the values with moderate accuracy although the 
 prediction accuracy for these properties is reasonably high for some datasets (INFOCOM 2006, SIGCOMM 2009).
 \item The prediction accuracy is poor across all datasets for betweenness centrality and in some cases for clustering coefficient and closeness centrality.
\end{itemize}

An important observation is that the spectrogram analysis (introduced in section ~\ref{properties}) is able to distinguish between these properties based on their 
 predictability. On ranking the properties based on the PSD value at bin 1 (refer to figures ~\ref{fig_all_dataset}(B), (D), (F) and ~\ref{fig_all_dataset_1}(B), (D)), we observe 
 that the higher ranked properties are the ones for which the prediction error is low while the lower ranked ones have higher prediction error. 
 Following this observation we further plot the mean percentage error for all the properties across all the datasets versus LPSD in figure ~\ref{k1}. 
 The plot clearly shows that the higher the value of LPSD, lower is the 
mean percentage of error and vice versa.

On further investigating into the cases where the prediction error is high, we observed that these points are mostly located either in places where a sharp 
 transition occurred or in silent phases where there was limited interaction among the nodes. Figure~\ref{fig9} identifies some of the transition and silent phases 
 in the time series of number of active edges in INFOCOM 2006 dataset.
Similar phases are also present in the other datasets as well.
\begin{figure*}[!ht]
  \begin{center}
  \includegraphics[width=0.9\linewidth, angle=0]{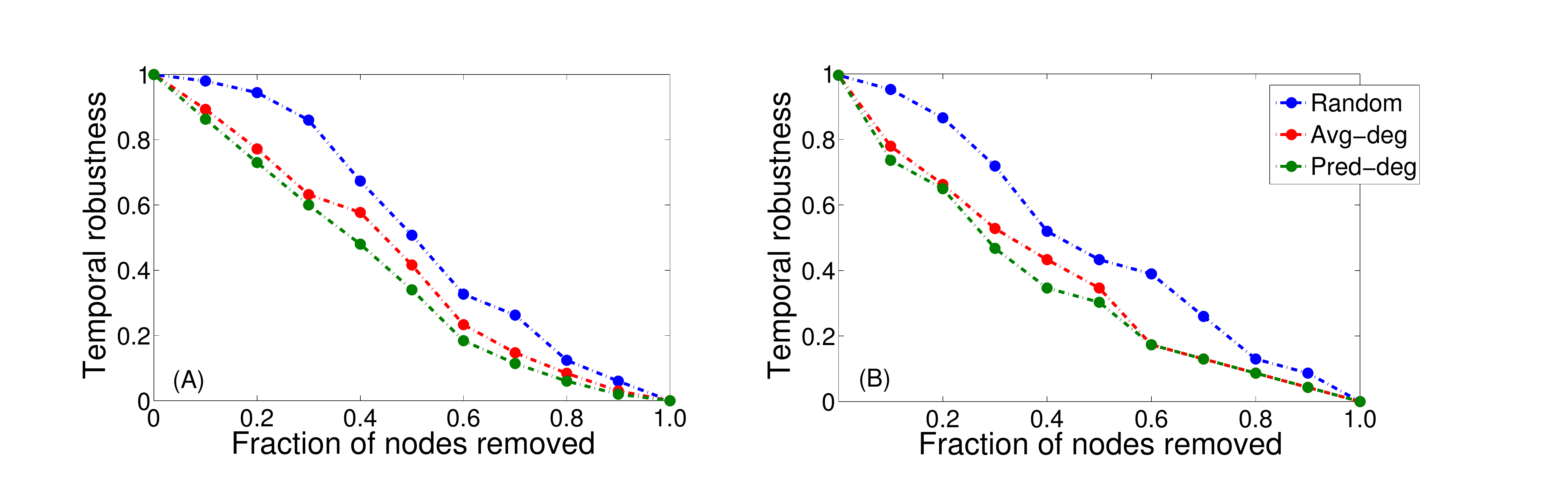}
  \caption{\label{fig_attack}Temporal robustness as a function of the fraction of removed nodes for (a) INFOCOM 2006 and (b) SIGCOMM 2009 datasets.}
  \end{center}
 \end{figure*}

\subsection{Enhancing the prediction scheme through spectrogram}

Since spectrogram analysis of the time series could determine the predictability of the corresponding property, 
an immediate extension would be to check whether it could be leveraged to identify beforehand the cases where the prediction error is high (unsuitable for prediction). 
To that aim we extend the spectrogram analysis to the single point case whereby  
 while predicting a property at a give time step, we find that the spectrogram of the window ($w$) and use the LPSD value as an indicator for potential prediction accuracy. 
 The cases identified by spectrogram analysis to be unsuitable for prediction can then be filtered out to improve the overall accuracy of the prediction framework. 
  A schematic diagram of this enhanced prediction framework is 
 provided in figure ~\ref{fig13}.
We now consider all the datasets and the corresponding time steps for prediction (which we considered earlier in this section, refer to table 1) and  
instead of directly using our prediction framework we 
 perform spectrogram analysis (single point method) on these points to separate out those which are unsuitable for prediction. We predict 
 only the points which the spectrogram analysis identified as suitable for prediction.
 In table ~\ref{tab_res} we compare the fraction of predictions with 
 error $\leq 20\%$ between both the cases where we do not use spectrogram  and where we use spectrogram. 
 We observe that the fraction of prediction with error $\leq 20\%$ is enhanced for all the properties albeit only marginally in some cases where the accuracy was already high. 
 More importantly for (ill-predicted) properties like betweenness centrality, closeness centrality, the prediction
accuracy increases substantially. 
 Note that prediction error $20\%$ is again representative and similar results 
 could be obtained for other values of prediction error as well.

\section{An attack strategy using prediction framework}
\label{attack}
In this section we show how our prediction can be used in order to launch targeted attack on temporal networks. The strategy proposed is a modification over the average node degree attack presented in ~\cite{trajanovski2012error}. 
In case of average node degree attack the temporal degree\footnote{Given a time interval [$t_1, t_n$] temporal degree of node $i$ is the average degree of $i$ over the time interval}  
~\cite{trajanovski2012error} of the nodes are calculated and the node with highest temporal degree is removed in the 
subsequent steps (i.e., ``the node is attacked''). 
We observe that for every 
node its degree over a given time interval forms a time series. Using our prediction framework we calculate the degree of the node at a future time step based on the previous 
$w$ time steps (window size for the corresponding dataset, refer to section \ref{prediction})
and remove a node with the 
highest degree as predicted by our proposed framework. We compare our strategy (Pred-deg) with average node degree based attack (Avg-deg) and the random case (nodes are selected at random 
and removed). 
The effectiveness of an attack strategy is measured using temporal robustness ~\cite{trajanovski2012error} which is estimated by the relative change in efficiency ~\cite{trajanovski2012error} 
after a structural damage $D$. 
Temporal efficiency of a network $G$ in a given time interval [$t_1,t_n$], $E_G(t_1,t_2)$ is defined as the averaged sum of the inverse temporal distances over all pairs of 
nodes in that time interval. 
\begin{center}
 $E_G(t_1,t_2)=\frac{1}{N(N-1)}\Sigma_{i,j:i\neq j}\frac{1}{d_{ij}(t_1,t_2)}$
\end{center}
Here $N$ is the number of nodes in the network and $d_{ij}(t_1,t_2)$ is the temporal distance which is the smallest temporal length paths among all the temporal
paths between $i$ and $j$ in the time interval [$t_1,t_2$]. Hence, temporal robustness is defined  as $R_G(D)=\frac{E_{GD}}{E_G}$.
In figure~\ref{fig_attack} we plot temporal robustness 
as a function of the fraction of nodes ($P$) removed 
for (a) INFOCOM 2006 and (b) SIGCOMM 2009 datasets. 
We observe that our strategy 
does better than both the random and average node degree based strategy.

\section{Conclusions and future work}
\label{conclusion}
Our contributions in this paper can be summarized as below:\newline
\textendash \ we provide a general framework to map temporal network of human contacts consisting of a series of graphlets equispaced in time into time series
and provide a detailed time domain and frequency domain analysis.\newline
\textendash \ we re-establish the presence of structural correlation in a temporal network of human face-to-face contact using a new metric which we call
neighborhood-overlap.\newline
\textendash \ we further quantify the extent of this correlation using neighborhood-overlap and use to identify the correct window used in our prediction framework.\newline
\textendash \ we also provide an approach for predicting the properties of future network instances using time series as a proxy and show that even though the 
precise network structure is not known at time step, one can estimate its properties.\newline
\textendash \ finally we provide a frequency domain analysis of temporal network and show how it can be useful in enhancing the prediction accuracy.\newline
\textendash \ as an application we show how our framework can be used in devising better strategies for targeted network attacks.\newline
 In its current state our framework can predict the values of the network properties at a future time step but is unable to offer the exact network 
 structure at that time step. But our framework can have genuine contributions toward link prediction in temporal networks. Since we show that structural 
 correlation exists in these networks and we can also predict the network properties at these time steps as well, we can re-frame the link prediction problem 
 as a network at a time step which is obtained from the network at a previous time step with minimal changes made depending on the values of the properties. 
 We plan to deeply investigate this problem in subsequent works.

 \bibliographystyle{unsrt}

\bibliography{ref}

\end{document}